\documentclass[a4paper,fleqn,usenatbib]{mnras}
\usepackage{amssymb,amsmath,graphicx,psfrag,mathtools}
\def\app#1#2{%
	\mathrel{%
		\setbox0=\hbox{$#1\approx$}%
		\setbox2=\hbox{%
			\rlap{\hbox{$#1\propto$}}%
			\lower1.1\ht0\box0%
			}%
			\raise0.25\ht2\box2%
			}%
			}

\usepackage[T1]{fontenc} 
\usepackage{ae,aecompl} 
\usepackage{url}
\usepackage{revsymb}
\hypersetup{draft} 
\title[Non-Repeating FRB]{The Sources of Apparently Non-Repeating FRB
}
\author[J. I. Katz]{
	J. I. Katz,$^{1}$\thanks{E-mail katz@wuphys.wustl.edu} 
\\
$^{1}$Department of Physics and McDonnell Center for the Space Sciences,
Washington University, St. Louis, Mo. 63130 USA 
}
\date{Accepted XXX.  Received YYY; in original form ZZZ} 
\pubyear{2022} 
\date{\today}
\begin{document} 
\label{firstpage} 
\pagerange{\pageref{firstpage}--\pageref{lastpage}} 
\maketitle 
\begin{abstract}
	There are insufficient catastrophic events (collapse, explosion
	or merger of stars or compact objects) to explain the cosmologically
	local rate of apparently non-repeating FRB if each such catastrophic
	event produces a single FRB.  Unless produced by some novel and
	unsuspected but comparatively frequent event, apparently
	non-repeating FRB must actually repeat many times in the lifetimes
	of their sources.  Yet no such infrequent repetitions (in contrast
	to the frequent activity of FRB known to repeat) have been observed,
	constraining their repetition rates and active lifetimes.  The
	absence of more frequent weaker but detectable repetitive outbursts
	in apparent non-repeaters resembles the distribution of SGR
	outbursts, with a large gap between giant outbursts and lesser
	outbursts.  This suggests mini-SGR as sources, more energetic than
	SGR 1935$+$2154 associated with FRB 200428 but less energetic than
	SGR 1806$-$20 that had no associated FRB.  Their largest radio
	outbursts would, at cosmological distances, be apparently
	non-repeating FRB and their X-ray and gamma-ray outbursts would be
	undetectable.  The large gap between the strongest outburst of FRB
	200428 and its lesser outbursts resembles the gamma-ray
	properties of individual well-observed SGR; at twenty times its
	actual distance, FRB 200428 would have been an apparent
	non-repeater. 
\end{abstract}
\section{Introduction}
I argue that repeating Fast Radio Bursts (FRB) and apparently non-repeating
FRB are fundamentally different objects.  Complementary to \citet{K22} about
repeating FRB, this paper advances a hypothesis for the origin of apparent
non-repeaters.  It suggests, on the basis of statistical similarities of
their distributions of strengths that they are more energetic relatives of
FRB 200428/SGR 1935$+$2154 but unrelated to known repeating FRB.

The measured volumetric rate of apparently non-repeating FRB in the local
Universe equals or exceeds the rates of all known catastrophic events:
collapses, explosions or mergers of compact objects, including white
dwarfs, neutron stars and black holes \citep{H20}.  The actual rate of
apparently non-repeating FRB must be greater than their measured rate,
likely by a large factor, because some, perhaps the overwhelming majority,
must be too faint to be detected as a result of beaming, insufficient
radiated energy or a combination of these.

It is implausible that the distribution of fluxes or fluences cuts off
at our detection threshold.  The intrinsic cutoff of radiated flux density
or energy implied if there is no undercount of more distant FRB (if all
are above the detection threshold at their distances) would imply a cutoff
in the flux or fluence distribution of closer FRB (where this emission
cutoff would occur at fluxes or fluences far above the detection threshold).
No such cutoff is observed.  The true event rate must far exceed the
observed rate.

This implies that apparently non-repeating FRB must either repeat or be
emitted by sources are not the product of catastrophic events.  Plausible
models associate them with ultra-magnetized neutron stars (``magnetars''),
the results of catastrophic stellar collapse, likely in supernov\ae, that
produce Soft Gamma Repeaters (SGR).  This paper attempts to reconcile the
the absence of a FRB associated with the giant outburst of SGR 1806$-$20
with the temporal and spatial association of FRB 200428 with lesser
outbursts of SGR1935$+$2154, and attributes apparently non-repeating FRB
to mini-SGR outbursts, perhaps of neutron stars with magnetic fields less
than those of magnetars.  This association is supported by similarities
between the statistics of outbursts from a single SGR source and those of
the bursts of FRB 200428.  Apparently non-repeating FRB are too distant for
their lesser bursts to be observed and their statistics measured.
\section{Empirical Bounds on Repetition Rates}
The definition of a non-catastropic burst model is one in which the {\it
mean\/} rate of repetitions does not vary substantially between bursts; a
single burst does not materially change the source's properties.  When
many possible sources may be in a field of view, the time required to
observe a burst in an initial survey is not an estimate of their repetition
rate.  Once a burst has been observed, its dispersion measure, which is
expected to be very stable, may permit unambiguous association with
subsequent bursts (the rotation measure, although observed to vary somewhat
in known repeaters, may be an additional source of discrimination).  A
quantitative test of the hypothesis of a constant rate of burst activity is
suggested in Appendix \ref{appen}.

Direct observational bounds on repetitions of apparently non-repeating FRB
are few.  \citet{R19} found no repetitions in 78 hours of observation of
FRB 190523 spread over 54 days, setting a $1\sigma$ upper limit on the
detection rate at their threshold of $\lesssim 0.3/$d.  FRB 190523 was only
15\% over this threshold, so no information about the distribution of event
sizes can be inferred from the absence of weaker bursts from this source.

\citet{B19} detected FRB 180924 in an 8.5 h ASKAP observing session.  It had
a fluence of 16 Jy-ms and a signal-to-noise ratio of 21, 2.1 times the
minimum ratio of 10 (a fluence of 8 Jy-ms) usually taken in FRB astronomy as
the threshold of reliable detection.  They report no repetitions (or
precursor bursts) in 720 hours of lower sensitivity observations with a
threshold of 28 Jy-ms and no repetitions in 11 hours of higher sensitivity
observations at Parkes with a threshold of 0.6 Jy-ms \footnote{In both
cases the fluence threshold has been scaled as the $1/2$ power of the burst
width, taken to be the 1.3 ms width of FRB 180924.}. 

When different observing systems have differing sensitivities, their
effective periods of observation must be corrected.  If the distribution
function of the source signal strength $F$ (flux or fluence) is $\propto
F^{-\alpha}$, the actual duration $T_2$ of the second observation is
replaced by an effective $T_2^\prime = T_2 (F_2/F_1)^{-\alpha}$, where
$F_2$ and $F_1$ are the detection thresholds, fluxes or fluences, of the
second and discovery observing systems respectively.

\citet{Z21} fitted $\alpha \approx 1.85$ to their entire dataset for the
repeating FRB 121102.  They also divided their data into two subsets,
finding $\alpha = 1.7$ in one and $\alpha = 2.6$ in the other.  The
applicability of these data to an apparently non-repeating FRB like FRB
180924 is necessarily uncertain.  With $\alpha = 1.85$ the $T_2 = 720\,$h of
lower sensitivity (single dish) observations of \citet{B19} correspond to
$T_2^\prime \approx 70\,$h while their 11 h of higher sensitivity (Parkes)
observations correspond to $T_2^\prime \approx 1500\,$h.  For the sensitive
Parkes observations $\alpha = 1.7$ would imply $T_2^\prime \approx 1000\,$h,
while $\alpha = 2.6$ would imply $T_2^\prime \sim 10^4\,$h.  The absence of
any detected bursts in these observations predicts with confidence $1 -
1/\beta$ a lower bound of $> T_2^\prime/\ln{\beta}$ on recurrence times of
FRB 180924 at the strength of its initial (and so far only) detected
outburst, or $\gtrsim 300\,$h with 95\% confidence.  Recurrence times orders
of magnitude longer (or infinite) are consistent with this limit, but not
with the assumption that repeaters and apparent non-repeaters differ only in
repetition rate.

The twice-daily revisits of the CHIME/FRB beam to points in its field of
view, with typical exposures $\approx 15\,$m, mean either that repetitions
(at comparable or greater strength) of FRB it observes will be seen or that
stringent upper bounds on their frequency will be established.  CHIME/FRB's
apparent non-repeaters can be distinguished {\it a priori\/} from repeaters
by some other property, such as ``sad trombone'' frequency drifts, then
the absence of repetition of $N \approx 500T/$y sources in an observing time
$T$ with duty factor $D\approx 30\,\text{m}/1\,\text{d} \approx 2 \times
10^{-2}$ sets an upper
limit to their repetition rate
\begin{equation}
	R \lesssim {1 \over NDT} \sim {0.1 \over \text{y}}\left({T \over
	\text{y}}\right)^{-2}
\end{equation}
\citep{CHIME20}.  One year of data can establish a repetition time $\gtrsim
10\,$y, while ten years may set a lower bound $\sim 10^3\,$y, or detect
repetitions if their characteristic rate is more than once per millenium.
This limit or value would be inferred from the repetition of a single source
or the absence of any repetitions in a database of many sources.
\section{Outliers}
A study \citep{K21} of the energy distributions of astronomical objects and
events showed that in almost all of them the most energetic object or event
is statistically consistent with extrapolation of the observed power law
distributions of lesser objects or events.  When this is true, it is
plausible that all the events are qualitatively similar, differing only in
some continuous variable, usually energy scale.

There were only two exceptions: the distribution of gamma-ray fluences of
outbursts of SGR 1806-20, and the distributions of fluxes and fluences of
outbursts of FRB 200428.  In both, the hypothesis of consistency could be
rejected at a confidence level $> 99.9\%$:  In the case of FRB 200428 the
power law exponent is uncertain because only three lesser bursts have been
observed \citep{Z20,Ki21}, but the outlier, the MJy-ms discovery burst, is
so extreme that its outlier nature cannot plausibly be doubted.  This is
entirely different from the statistics of known repeating FRB, whose
brightest outbursts are consistent with the power laws fitted to their
lesser outbursts.

This statistical observation is an argument for identifying objects like FRB
200428 with ``magnetar'' SGR, independent of its observed coincidence,
temporal as well as spatial, with SGR 1935$+$2154.  If FRB 200428 had been
beyond 100 kpc (and within $\sim 12\,$Mpc, so that its major outburst would
have been detectable; \citet{CHIME20}), it would have been classified as an
apparently non-repeating FRB.
\section{Models}
\citet{K22} attributed known repeating FRB with detected event rates as high
as $\sim 10^{-2}$/s (varying as a negative power of the detection threshold,
again implying the absence of an intrinsic energy scale) to accretion discs
or funnels around intermediate mass black holes.  Any outbursts of such
systems have no natural energy scale, and therefore must have a power law
distribution, as in models of ``self-organized criticality''
\citep{K86,BTW87}.  Events without a characteristic scale must have a
power law distribution because a break in the power law would define a
characteristic scale, and conversely.

The absence of detected
weaker repetitions of apparently non-repeating FRB suggests an association
with SGR, the frequency of whose giant outbursts is not described by an
extrapolation of the frequency distribution of lesser outbursts; the giant
outbursts are true outliers, qualitatively different from the lesser events
\citep{K21}.  This association is supported by the geometrical and temporal
coincidence of FRB 200428 with an outburst of SGR 1935$+$2154
\citep{B20,CHIME20,L20,M20,R20,T20}.  The fact that the few detected
repetitions of FRB 200428 were four \citep{Ki21} to eight \citep{Z20} orders
of magnitude fainter than its April 28, 2020 MJy-ms burst is consistent with
the distribution of SGR outbursts
\citep{G00,H05,P05,G06,G07} but not with those of other repeating FRB.
\section{Discussion}
No FRB was associated with the giant outburst of SGR 1806-20 \citep{TKP16}.
This is explicable because such giant outbursts fill the circumstellar space
with opaque equilibrium pair plasma \citep{K96} that would preclude the
acceleration of relativistic particles required to emit FRB as well as the
escape of coherent radio radiation.  In equilibrium, a region the size of a
neutron star fills with opaque pair plasma if $k_B T \ge 22\,$keV, a value
only logarithmically dependent on the plasma dimensions.  If such a plasma
covers the surface of a neutron star its soft gamma-ray luminosity is
$\gtrsim 10^{42}\,$erg/s and expected to have a thermal spectrum.  If the
energy density is less than that corresponding to pair-black body
equilibrium at 22 keV, equilibrium may not be achieved.  A localized flare
may also be smaller than the neutron star surface area.  This is consistent
with the non-thermal X-ray spectrum and peak power of $10^{40}$--$10^{41}$
erg/s of SGR 1935$+$2154 during the giant radio burst of FRB 200428
\citep{L20,M20,R20,T20}.

This paper predicts that SGR outbursts with luminosity $\gtrsim
10^{42}\,$erg/s (like the giant outburst of SGR 1806$-$20) do not emit FRB
while less luminous outbursts may, but do not necessarily, emit them.
Apparently non-repeating FRB must be the products of less energetic
``mini-SGR'', although to be detectable at cosmological distances their
associated FRB must be orders of magnitude more energetic than FRB 200428.

It will be necessary to distinguish rarely repeating ``true apparent
non-repeaters'' from ``repeaters''.  This may be possible on the basis of
the distribution of intervals between bursts with fluences within moderate
(2--10) factors of each other and the ratios of brightest to
second-brightest bursts.  These ratios are ${\cal O}(1)$ in known repeaters
but $\ggg 1$ in FRB 200428.  The hypothesis of this paper and of the
complementary \cite{K22} is that these ratios are bimodal.
\section*{Acknowledgment}
I thank V. Kaspi for explaining the capabilities of CHIME/FRB.
\section*{Data Availability}
This theoretical study did not generate any new data.

\appendix
\section{Testing the Hypothesis of Constant Rates}
\label{appen}
If the direction of FRB 180924 had been identified before it was observed
(for example, by the presence of some object of interest) the hypothesis of
a constant mean repetition rate could be tested if sufficient data existed.
This was not the case; it was observed in a single 8.5 hour observing run
that was part of 12,000 hours of observation \citep{B19}.  The following
procedure might be useful when {\it a prior\/} knowledge of the object
exists.  It is not useful here because the effective observing time $T_1$ is
the entire 12,000 hours of the survey; to show its potential utility the
inapplicable $T_1 = 8.5\,$h of the single observing run in which FRB 180924i
was observed is used in the numerical estimates.

If one and only one burst is observed in an observing time $T_1$, repetition
rates $r < 1/(\beta T_1)$ for $\beta \gg 1$ may be excluded with confidence
$\approx 1-1/\beta$.  The probability of no burst in an independent
observation of duration $T_2$ is then $\exp{(-rT_2)} <
\exp{[\beta(T_2/T_1)]}$.  The absence of a burst would reject the constant
repetition rate hypothesis at a level of confidence $\gamma = 1 -
\exp{(-rT_2)}$.  Setting $\gamma = \beta$ indicates that the duration of
observation required to reject the constant rate hypothesis is
\begin{equation}
	T_2 = T_1 \beta \ln{\beta}.
\end{equation}
Rejection at the 95\% confidence level ($\beta = 20$) requires $T_2 > 60
T_1$.

When different observing systems have differing sensitivities, $T_2$ is
replaced by an effective $T_2^\prime = T_2 (F_2/F_1)^{-\alpha}$, where
$F_2$ and $F_1$ are the detection thresholds, fluxes or fluences, of the two
systems and the distribution function of the source signals is assumed to be
$\propto F^{-\alpha}$; for FRB $\alpha \approx 1.85$ \citep{Z21} is
plausible, although it appears to be variable.  With this scaling, the
720 hours of lower sensitivity observations of \citet{B19} have $T_2^\prime
\approx 70\,$h while their 11 h of higher sensitivity observations have
$T_2^\prime \approx 1500\,$h, implying $T_2^\prime/T_1 \approx 170$.  This
is sufficient, given the assumptions (most significantly, the energy
distribution of the bursts), to exclude the constant rate hypothesis with
a confidence level $> 97\%$.  \cite{Z21} suggest $\alpha = 1.7$ in one
subset of their data and $\alpha  = 2.6$ in another subset.  The lower
value would imply (for the Parkes data) $T_2^\prime \approx 1000\,$h,
$T_2^\prime/T_1 \approx 115$ and rejection of the constant rate hypothesis
at a confidence level $\approx 97\%$; the higher value of $\alpha$ would
increase $T_2^\prime/T_1$ to $10^4$ and rejection to a confidence level $>
99\%$\footnote{Let us not forget Ehrenfest's admonition \citep{G63} ``If it
is essential to use probability to prove that you are right, you are usually
wrong.''}.
\label{lastpage}

\begin{thebibliography}{99}
\bibitem[\protect\citeauthoryear{Goudsmit}{1963}]{G63} Goudsmit,
		S. A. 1963 \prl\ 10, 149 This editorial comment is not
		indexed and is not in the ADS database.  It may be found in
		hard copies.
\end{thebibliography}
\end{document}